# Control of InGaAs facets using metal modulation epitaxy (MME)

Mark A. Wistey[a]

Electrical Engineering Department, University of Notre Dame, Notre Dame, Indiana 46656, USA

Ashish K. Baraskar

GLOBALFOUNDRIES, Malta, New York, 12020, USA

Uttam Singisetti

Electrical and Engineering Department, University at Buffalo, Buffalo, New York 14260, USA

Greg J. Burek

Department of ECE, University of California, Santa Barbara, California 93106, USA

Byungha Shin, Eunji Kim, Paul C. McIntyre

Materials Department, Stanford University, Stanford, CA 94305, USA

Arthur C. Gossard, Mark J. W. Rodwell

Departments of Materials and ECE, University of California, Santa Barbara, California 93106, USA

[a]Electronic mail: mwistey@nd.edu

Control of faceting during epitaxy is critical for nanoscale devices. This work identifies the origins of gaps and different facets during regrowth of InGaAs adjacent to patterned features. Molecular beam epitaxy (MBE) near $SiO_2$ or $SiN_x$ led to gaps, roughness, or polycrystalline growth, but metal modulated epitaxy (MME) produced smooth and gap-free "rising tide" (001) growth filling up to the mask. The resulting self-aligned FETs were dominated by FET channel resistance rather than source-drain access resistance.

Higher As fluxes led first to conformal growth, then pronounced {111} facets sloping up away from the mask.

## I. INTRODUCTION

Nanoscale devices have many advantages, including high bandwidth and packing density. But the gate oxide in MOSFETs has become difficult to shrink, leading to short-channel effects and off-state leakage currents. Further improvement in FET performance could come from semiconductors with higher carrier velocities. InGaAs and other III-V materials have electron velocities 5-10 times higher than those in silicon, producing strong interest in III-V MOSFETs.[1-5] Significant progress has been made on III-V dielectrics and interface control layers, and scalable CMOS-like process flows have been demonstrated.[6] But high source/drain resistances have hindered device performance. Contacts are challenging because III-V's lack an equivalent to the highly conductive salicides used for Si CMOS, though reacted contacts[7] and NiInAs[8] have shown contact resistances below $10^{-8}$ $\Omega$-cm$^{-2}$.

Source/drain resistance also results from heterojunction barriers, long distances, and low carrier densities, in addition to contact resistance.[9] Even with doped channels for depletion-mode FETs, the typical distances between metal and device introduce parasitic access resistance, which impairs high frequency operation.[10] Making the contacts self-aligned would greatly reduce access resistance without requiring critical lithographic alignment.[11] Self-aligned dopant implants in III-V's lack the necessary active carrier concentrations (above $2\times10^{19}$ cm$^{-3}$) to prevent source exhaustion in thin channels at CMOS current densities.[12-14]

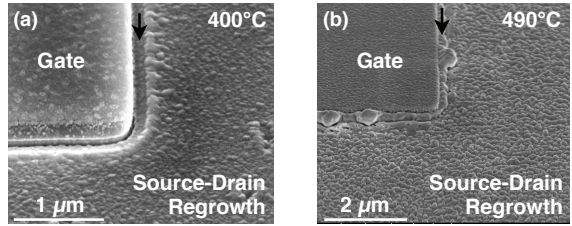

FIG. 1. Top view scanning electron micrograph (SEM) of MBE regrowth near a SiO$_2$-masked gate. Note 200 nm gap in regrowth near mask at low growth temperatures near 400 °C (a), and polycrystalline growth at ≥490 °C (b). .

We previously demonstrated regrowth of highly doped, InGaAs contacts by molecular beam epitaxy (MBE), but these showed gaps or tapered regrowth near gate masks, as shown in Fig. 1. The resulting access resistance was unacceptably high, leading to poor MOSFET performance.[15] Also, shorter gate lengths (narrower mesas) with $L_g$<500 nm produced slightly smaller gaps in regrowth. Similar self-aligned MBE InGaAs regrowth for tunnel FETs showed difficulty in the control of facets near the mask, and an unexplained moat or gap was apparent near several devices, wider than the gate overhang, similarly reported by Chun.[16]

One possible explanation for this difficulty is a local change in the ratio of Group V to Group III atom species on the surface. The III/V ratio can greatly affect surface kinetics during epitaxial growth and promote formation of different facets near step edges or raised features such as FET gates. Shen and Nishinaga reported from microprobe

RHEED analysis that for all InAs growth temperatures, increased arsenic flux led to faster growth on the (111)A plane, producing a flat (001) surface,[17] but the reverse was true near (111)B facets on GaAs.[18]

## II. EXPERIMENTAL

This work examines the origin of gaps and roughening in the regrowth of InGaAs near dielectric features, specifically a $SiO_2$ mask, with or without $SiN_x$ sidewalls, that fully encapsulated a FET metal gate. Two sets of samples were patterned on InGaAs lattice matched to InP, then verified by fabricating FETs. Transmission length methods (TLMs) far from device features did not accurately measure resistance of regrown InGaAs near FET gates,[19] so all samples in this work used a FET-like geometry.

All regrowths were performed in an Intevac Mod Gen II MBE using a valved arsenic cracker. Growth temperatures were measured using a Modline 3V pyrometer calibrated by band edge thermometry. Before regrowth, each patterned sample was exposed to UV ozone for 20 minutes to remove trace organics and form a sacrificial oxide. It was then dipped in 1:10 $HCl:H_2O$ for 60 seconds, followed by a 60 second rinse in deionized water. The wafer was immediately loaded into ultrahigh vacuum (UHV) and baked at 200 °C overnight. The wafer was then exposed to thermally cracked $H_2$ at $1\times10^{-6}$ Torr for 30 minutes at 420 °C as measured by non-contact thermocouple, with occasional rotation to assure uniform exposure of H from various angles. Reflection high energy electron diffraction (RHEED) showed a clear (2×4) reconstruction at 200 °C before the regrowth began, indicating a nominally clean surface.

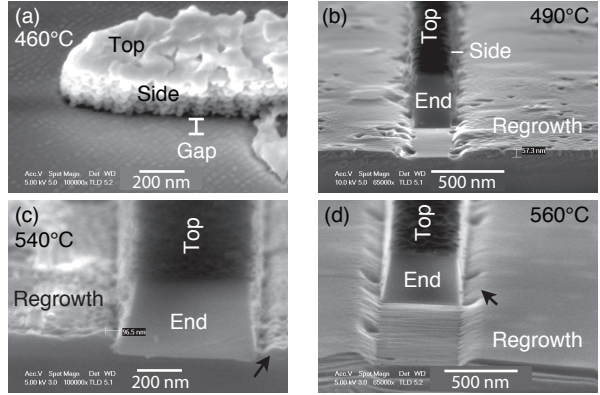

FIG. 2. Oblique side view (a) and cleaved face end view (b-d) SEM of MEE regrowth near $SiO_2$ mask (dummy gates) at different growth temperatures. Above 490 °C, regrowth showed no gaps and fewer pinholes, but facets (arrows) persisted near mask.

The first set used simple $SiO_2$ masks (dummy gates) followed by migration enhanced epitaxy (MEE)[20-25] for source/drain regrowth, to attempt to fill the gaps observed in Fig. 1. $SiO_2$ masks were patterned by photolithography, then the wafer was cleaned as above and loaded for regrowth. Group III fluxes were In=$9.7\times10^{-8}$ and Ga=$5.1\times10^{-8}$ Torr for $T_{sub}$ < 540 °C. Above $T_{sub}$ > 540 °C, In fluxes were increased to compensate for In desorption, calibrated by x-ray diffraction. As shown in Fig. 2, MEE growth quality improved at higher temperatures, with no gaps near masks, fewer pinholes, and less crosshatching. But facets persisted near masks, and access resistance was very high.

The second set of samples used metal modulation epitaxy (MME, Fig. 3)[26] to force longer and more uniform surface migration regardless of distance from mask. MME is similar to periodic supply epitaxy (PSE)[27] but with lower As flux to ensure high Group

III surface mobility. To best reproduce the geometry, strain, and other local conditions near actual FETs, this set used a complete MOSFET gate stack as detailed in Ref. 15, including $Al_2O_3$ high-$k$ dielectric and metal gate. The metal was covered by patterned $SiO_2$ and encapsulated in conformal 20-30 nm $SiN_x$ sidewalls. The $Al_2O_3$ high-$k$ was etched by dilute KOH, exposing the InGaAs surface for regrowth, leaving the $SiO_2$, $SiN_x$, and newly-exposed InGaAs intact. The processed wafers were then cleaned and loaded for regrowth as above.

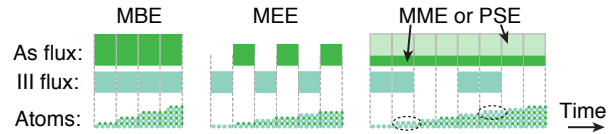

FIG. 3. Typical flux timing diagrams for MBE, MEE, MME, and PSE, and cross section of atom arrangements. Squares represent ~1 monolayer. Flux ratio V/III>>1 in MBE and PSE, ~1 in MEE, and <1 in MME. III-on-III (circled) has very high surface mobility.

The MME consisted of Group III deposition for 3.8 seconds to grow approximately 2 monolayers of InGaAs, followed by a 15 second pause under the same constant $As_2$ flux. This cycle was repeated 80 times to grow 40 nm of InGaAs. Unlike traditional MEE at low temperature, the arsenic flux was not interrupted, since InGaAs would decompose at these temperatures. Total InGaAs Group III fluxes (beam equivalent pressure) were $1.5\times10^{-7}$ Torr. Silicon doping was provided simultaneous with each Group III pulse, corresponding to a doping level of $[Si]=8\times10^{19}$ $cm^{-3}$ and $n=5\times10^{19}$ $cm^{-3}$.

Arsenic fluxes were $5.6\times10^{-7}$, $1.0\times10^{-6}$, $2\times10^{-6}$, and $5\times10^{-6}$ Torr for the respective InGaAs layers, ending with conditions similar to those in Fig. 2(d). Marker layers of 20 nm InAlAs were grown by conventional MBE with an $As_2$ flux of $5\times10^{-6}$ Torr. Substrate temperatures were decreased from 540 °C to 500 °C during the InAlAs layers, ensuring a smooth and conformal surface and freezing the surface profile of the underlying InGaAs for later analysis. No extra pauses were used before or after the InAlAs, in order to prevent surface profile changes from annealing. The InAlAs was also doped with $[Si]=8\times10^{19}$ $cm^{-3}$.

RHEED showed a continuous (4×2) pattern during the first three InGaAs layers, indicating a Group III-rich surface. It did not change during the arsenic-only steps, nor from one InGaAs layer to the next, although it did revert to a conventional (2×1) pattern during the InAlAs layers. RHEED during the fourth InGaAs layer, with highest As flux, oscillated between Group III rich (4×2) during the Group III pulses and Group V rich β2(2×4) during the pauses. RHEED during the InAlAs layers was initially spotty with substrate temperatures near 540 °C, but it became streaky again as substrate temperatures approached 500 °C, indicating a smooth surface.

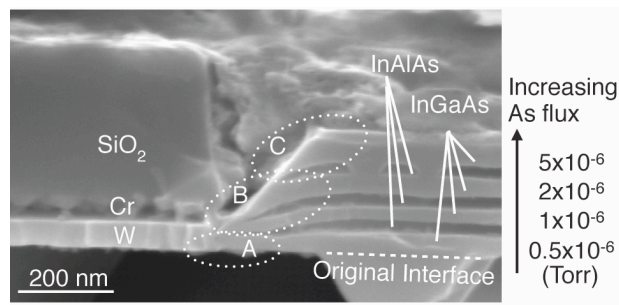

FIG. 4. InGaAs:Si layers grown with increasing As fluxes, separated by InAlAs marker layers. A: Lowest

arsenic flux shows "rising tide fill" without gaps near gate or $SiO_2/SiN_x$. B: Conformal growth. C: Complete {111} faceting. The conformal $SiN_x$ sidewall over the $SiO_2$, Cr, and W is not visible at this resolution.

Fig. 4 shows a scanning electron microscopy (SEM) cross section of the growth. A brief stain etch using dilute HCl was used to distinguish InGaAs from InAlAs. InGaAs with the lowest $As_2$ flux ($5\times10^{-7}$ Torr) filled the entire (001) plane right up to the mask. Higher As fluxes produced a tapered surface, with no further fill along the gate sidewall. The highest $As_2$ fluxes ($5\times10^{-6}$ Torr) produced growth terminated by {111} planes sloping up away from the mask. InAlAs layers also showed some thinning next to the mask due to shadowing of source material by the tall gate stack and off-normal MBE cell geometry. There was no visible pileup near the (001)/{111} step edges, which indicates high Ga/In surface mobility on the (001). There was no visible selectivity between (111)A and (111)B surfaces, which we interpret as an indication of fully Group III rich surfaces. We observed no significant differences in facet angles for masks aligned along (110) or ($\bar{1}$10). A constant average RHEED intensity suggested there was no Ga droplet formation, and no droplets were visible in SEM.

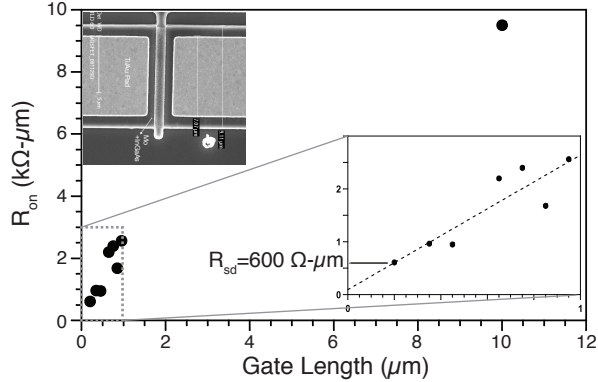

FIG. 5. Total on-state resistance vs. gate length (mask finger width) $L_g$ for FETs with regrown InAs contacts. Dashed line is fit over 0.3-1.0 μm. Inset: Completed FET with Ti/Au source/drain pads straddling a $L_g$=1 μm mask (gate) finger.

To verify these results and also test them with InAs, which makes low resistance n-type contacts, we fabricated actual FETs. Source/drain regrowth of a single 50 nm layer of either InGaAs or relaxed InAs was done by MME using As=$5\times10^{-7}$ Torr, then capped with in-situ molybdenum and processed into FETs as in Ref. 15. SEM and transmission electron microscopy (TEM, not shown) showed good filling next to the mask for both InGaAs and relaxed InAs, and little growth on sidewalls. On-state resistance vs. gate length is plotted in Fig. 5. Unlike earlier devices, the access resistance (extrapolation to $L_g$=0) was now a small fraction of total on-state resistance. The smallest gate lengths available were 300 nm for the InGaAs regrowth and 200 nm for the InAs regrowth. Total on-state resistivity Rsd was 600 Ω-μm for $L_g$=200 nm, so source and drain resistances are, at most, 300 Ω-μm each, and likely much lower.

## III. RESULTS AND DISCUSSION

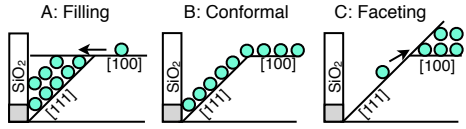

FIG. 6. Surface mobility affects facet competition. Planes with high surface mobility suffer a net loss of atoms to planes with stronger bonds. A-C correspond to conditions observed in Fig. 4.

We interpret these results as follows. Facet competition occurs when adatoms can move from one facet to another, as shown in Fig. 6. The facet with higher surface mobility generally has weaker bonds and loses atoms to its neighbor. From another perspective, the residence time for Group III adatoms is longer on a slow-diffusion surface, providing more time for additional atoms to arrive and bond them in place. Thus the facet with low surface mobility grows thicker but not wider. The facet with high surface mobility gets wider but not thicker as atoms move to a neighboring facet.[28] Smooth, low-index facets suggest a negligible or negative Schwoebel barrier.

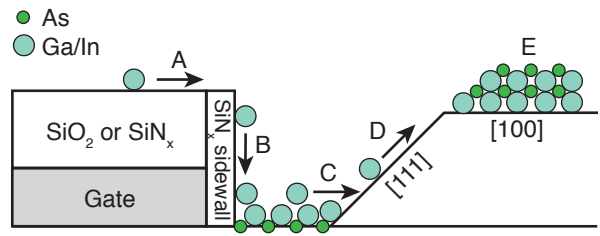

FIG. 7. Gap formation mechanism. A: Weak adhesion of In and Ga on $SiO_2$ produces migration to neighboring semiconductor. B: Arriving Ga/In atoms locally enrich the III/V ratio (e.g. Ga on Ga), leading

to rapid surface diffusion (C). As III/V ratio decreases farther from mask, growth begins on low-index facets (D) and eventually becomes planar (100) and As-terminated (E).

The gaps in regrowth next to surface features can be explained by two separate effects, both based on local changes in the III/V ratio. First, the incorporation mechanisms of As and Ga/In are different. In and Ga tend to migrate on the growth surface, while As tends to evaporate and be replaced.[29] During growth, tall features (gates) block some As flux from surrounding areas, while Ga and In continue to migrate until reaching areas with higher As flux. Second, at growth temperatures much above 400 °C, In and Ga tend to bond relatively weakly on $SiO_2$ or $SiN_x$, so they can readily migrate to neighboring semiconductor. As with shadowing, the III/V ratio is increased locally, as shown in Fig. 7. This mechanism explains the sensitivity to gate length (mask width), since a larger mask area can provide more Ga and In atoms, up to the limit of the surface diffusion length of Ga and In on the dielectric.

Low-As MME prevented gaps and {111} faceting by strongly increasing the Group III surface coverage, and therefore surface mobility, at all distances from the mask. As shown in Fig. 3, MME alternates between strongly As-rich and III-rich conditions. The far field no longer acted as a sink for Group III adatoms since diffusion rates were similar everywhere.

Shen and Nishinaga reported that decreased As flux led to migration of atoms from the (001) surface to (111)A planes during InAs growth.[17] This led to the (001) plane growing wider and {111} planes becoming less pronounced. The opposite was reported for GaAs near (111)B facets.[18] In contrast, we find that an increased arsenic flux increased faceting of both (111)A and (111)B, and the best gap-free fill next to dielectric-coated features occurred with the lowest arsenic flux. We note there are multiple differences between our growth conditions and Shen's, such as higher growth temperatures, Group III-rich pauses for migration enhancement, high Si doping, and $As_2$ rather than $As_4$.

We did not observe cusps in the regrowth. We attribute this to sufficiently high surface mobility on the (001) surface under all conditions, so adatoms did not pile up near the {111}-(001) intersections but instead diffused uniformly over the surface. Hata reported that Ga has a surface diffusion length of 1-8 µm on GaAs at somewhat higher temperatures (560 °C),[30] and In has an even higher surface mobility than Ga. The lack of a visible cusp sets a lower bound on (001) surface mobility of about 3 µm.

Nucleation and growth on the mask, visible in Fig. 2(a), could change local growth conditions over the course of the growth. The first InGaAs layer could have excess Group III atoms migrating from the $SiO_2$ cap to the semiconductor surface, but once nucleated, InGaAs on top of the mask would consume Group III. However, previously reported devices showed faceting with or without selective growth.[31] Although Fig. 4 shows growth on the sides of the mask, other samples did not, yet they all showed similar faceting next to masks.

The total on-state resistivity places an upper bound of 300 Ω-µm on source and drain resistivities. Actual resistivities are likely much lower than this, but scatter in the data precludes a confident extrapolation to $L_g$=0. Even so, this on-state resistance is an order of magnitude better than our previous enhancement-mode MOSFETs.

Finally, we note that the thinning of InAlAs near the mask is insufficient to explain InGaAs faceting. Single layers of InGaAs grown without InAlAs showed the same faceting under similar stoichiometry, such as in Fig. 2(d) and "C" in Fig. 4. Also, Fig. 4 clearly shows conformal, non-faceting InGaAs for [As]=$10^{-6}$ Torr even though the underlying InAlAs already has a wedge profile.

## IV. SUMMARY AND CONCLUSIONS

We found that varying growth temperature in both MBE and MEE of InGaAs on InP was insufficient to provide flat, high quality surfaces without gaps near dielectric masks including $SiO_2$ and $SiN_x$. Low temperatures left gaps, attributed to a local enhancement of the III/V ratio due to migration of In and Ga from the mask, and possibly shadowing of As by tall features (e.g. FET gate) during growth at lower temperatures. High growth temperatures created rough and defective material near the mask, possibly due to differences in surface mobility of Ga vs. In atoms leading to In-rich growth and strain relaxation.

On the other hand, metal modulation epitaxy (MME) enabled uniform surface mobility and homogeneous growth across the whole wafer, including areas near dielectric masks. Pulses of 2 monolayers of Group III atoms were grown under metal-rich conditions, followed by an As soak to consume the excess Group III atoms. MME eliminated gaps and pinholes and enabled self-aligned regrowth with no crosshatching.

Varying the As flux in MME also allowed control of the facets adjacent to the dielectric features (i.e. masks). High As fluxes produced well-defined {111} planes.

Fluxes closer to stoichiometry, marked by alternating (2×4) and (4×2) RHEED patterns with each growth cycle, led to conformal growth. Such facet control is important for self-aligned contacts and nanoscale self-assembled devices. Finally, a gap-free (001) "rising tide" fill along the mask was achieved when the As flux was roughly half that necessary to produce alternating RHEED patterns. MOSFETs with MME regrown InAs source/drain were not limited by access resistance, which was below 300 $\Omega$-$\mu$m.

.

## ACKNOWLEDGMENTS

The authors thank C. J. Palmstrøm for helpful discussions, and acknowledge the support of the Semiconductor Research Corporation (SRC) through the Non-Classical CMOS Research Center. A portion of this work was performed in the UCSB nanofabrication facility, part of the NSF funded NNIN network. This work made use of MRL Central Facilities supported by the MRSEC Program of the National Science Foundation under award No. MR05-20415.

[1] Y. Xuan, Y. Q. Wu, P. D. Ye, IEEE Electron Device Lett., **29**, 294 (2008).

[2] A. Delabie, D. P. Brunco, T. Conard, P. Favia, H. Bender, A. Franquet, S. Sioncke, W. Vandervorst, S. Van Elshocht, M. Heyns, M. Meuris, E. Kim, P. C. McIntyre, K. C. Saraswat, J. M. LeBeau, J. Cagnon, S. Stemmer, and W. Tsai, J. Electrochem. Soc. **155**, H937 (2008).

[3] C.-W. Cheng and E. A. Fitzgerald, Appl. Phys. Lett., **93**, 031902 (2008).

[4] R. J. W. Hill, R. Droopad, D. A. J. Moran, X. Li, H. Zhou, D. Macintyre, S. Thoms, O. Ignatova, A. Asenov, K. Rajagopalan, P. Fejes, I. G. Thayne and M. Passlack, Electron. Lett., **44**, (2008).

[5] T. D. Lin, H. C. Chiu, P. Chang, L. T. Tung, C. P. Chen, M. Hong, J. Kwo, W. Tsai, Y. C. Wang, Appl. Phys. Lett., **93**, 033516 (2008).

[6] J. Mo, E. Lind, L.-E. Wernersson, IEEE Electron Dev. Lett. **35**, 515 (2014).

[7] R. Dormaier, Q. Zhang, Y. C. Chou, M. D. Lange, Y. M. Yang, A. Oki, and S. E. Mohney, J. Vac. Sci. Technol. B, **27**, 2145 (2009).

[8] R. Oxland, S. W. Chang, Xu Li, S. W. Wang, G. Radhakrishnan, W. Priyantha, M. J. H. van Dal, C. H. Hsieh, G. Vellianitis, G. Doornbos, K. Bhuwalka, B. Duriez, I. Thayne, R. Droopad, M. Passlack, C. H. Diaz and Y. C. Sun, IEEE Electron Device Lett. **33**, 501 (2012).

[9] K. Dae-Hyun and J. A. del Alamo, International Electron Devices Meeting (IEDM 2006).

[10] N. Neophytou, T. Rakshit, and M. S. Lundstrom, IEEE Trans. Electron Devices, **56**, 1377 (2009).

[11] R. T. P. Lee, R. J. W. Hill, W.-Y. Loh, R.-H. Baek, S. Deora, K. Matthews, C. Huffman, K. Majumdar, T. Michalak, C. Borst, P. Y. Hung, C.-H. Chen, J.-H. Yum, T.-W. Kim, C. Y. Kang, Wei-E. Wang, D.-H. Kim, C. Hobbs, P. D. Kirsch, 2013 IEEE Intl. Electron Devices Meeting (IEDM 2013). doi: 10.1109/IEDM.2013.6724546

[12] C.P. Chen, T.D. Lin, Y.J. Lee, Y.C. Chang, M. Hong, J. Kwo, Solid-State Electron., **52**, 1615 (2008).

[13] Y. Xuan, Y. Q. Wu, and P. D. Ye, IEEE Electron Device Lett., **29**, 294 (2008).

[14] M. Passlack, P. Zurcher, K. Rajagopalan, R. Droopad, J. Abrokwah, M. Tutt, Y. B. Park, E. Johnson, O. Hartin, A. Zlotnicka, P. Fejes, R. J. W. Hill, D. A. J. Moran, X. Li, H. Zhou, D. Macintyre, S. Thorns, A. Asenov, K. Kalna, and I. G. Thayne, 2007 IEEE Intl. Electron Devices Meeting (IEDM 2007), 621 (2007). doi: 10.1109/IEDM.2007.4419016

[15] U. Singisetti, M. A. Wistey, G. J. Burek, E. Arkun, A. K. Baraskar, Y. Sun, E. W. Kiewra, B. J. Thibeault, A. C. Gossard, C. J. Palmstrøm, and M. J. W. Rodwell, Phys. Status Solidi C **6**, 1394 (2009).

[16] Y. J. Chun, T. Uemura and T. Baba, Jpn. J. Appl. Phys., **39**, L1273 (2000).

[17] X.Q. Shen, T. Nishinaga, J. Cryst. Growth **146**, 374 (1995).

[18] X.Q. Shen, D. Kishimoto, T. Nishinaga, Jpn. J. Appl. Phys. **33**, 11 (1994).

[19] U. Singisetti, M. A. Wistey, G. J. Burek, A. K. Baraskar, J. Cagnon, B. J. Thibeault, S. Stemmer, A. C. Gossard, and M. J. W. Rodwell, E. Kim, B. Shin, P. C McIntyre, Y.-J. Lee, Indium Phosphide & Related Materials, 2009 (IPRM 2009). URL: http://www.ece.ucsb.edu/Faculty/rodwell/publications_and_presentations/publications/2009_5_may_singisetti_IPRM_digest.pdf

[20] J. Zhang, J. H. Neave, B. A. Joyce, A. G. Taylor, S. R. Armstrong, M. E. Pemble, Appl. Surf. Sci. **60/61**, 215 (1992).

[21] Y. Horikoshi, M. Kawashima, H. Yamaguchi, Jpn. J. Appl. Phys., Part 2, **25**, L868 (1986).

[22] Y. Horikoshi, M. Kawashima, H. Yamaguchi, Jpn. J. Appl. Phys., Part 1, **27**, 169 (1988).

[23] F. Briones L. Gonzalez, M. Recio, M. Vazquez, Jpn. J. Appl. Phys., Part 2, **26**, L1125 (1987).

[24] J.R. Arthur, J. Appl. Phys. **39**, 4032 (1968).

[25] C. T. Foxon, B. A. Joyce, Surf. Sci., **50**, 434 (1975).

[26] S. D. Burnham, G. Namkoong, D. C. Look, B. Clafin, and W. A. Doolittle, J. Appl. Phys. **104**, 024902 (2008).

[27] G. Bacchin, T. Nishinaga, J. Cryst. Growth, **198/199**, 1130 (1999).

[28] P. Atkinson, D.A. Ritchie, J. Cryst. Growth **278**, 482 (2005).

[29] C.T. Foxon and B.A. Joyce, Surf. Sci. **64**, 293 (1977).

[30] M. Hata, A. Watanabe, and T. Isu, J. Cryst. Growth, **111**, 83 (1991).

[31] M. Wistey, G. Burek, U. Singisetti, A. Nelson, B. Thibeault, J. Cagnon, S. Stemmer, A. Gossard, M. Rodwell, S. Bank, Intl. Conf. Molecular Beam Epitaxy, Vancouver, British Columbia, Canada, (2008).

**Figure Captions**

Figure 1. Top view scanning electron micrograph (SEM) of MBE regrowth near a $SiO_2$-masked gate. Note 200 nm gap in regrowth near gate at low growth temperature (≈400 °C) and polycrystalline growth at ≥490 °C (respective arrows).

Figure 2. Oblique side (a) and cleaved face (b-d) SEM views of MEE regrowth near $SiO_2$ dummy gates, at different growth temperatures. Above 490 °C, regrowth showed no gaps and fewer pinholes, but facets (arrows) persisted near gates.

Figure 3. Typical flux timing diagrams for MBE, MEE, and MME, and schematic cross section of atom arrangements. Each square represents ~1 ML. V/III>>1 in MBE, ~1 in MEE, and V/III<1 in MME. III-on-III (circled) has very high surface mobility.

Figure 4. InGaAs:Si layers grown with increasing As fluxes, separated by InAlAs marker layers. A: Lowest arsenic flux shows "rising tide fill" without gaps near gate or $SiO_2/SiN_x$. B: Conformal growth. C: Complete {111} faceting. The conformal $SiN_x$ sidewall over the $SiO_2$, Cr, and W is not visible at this resolution.

Figure 5. Total on-state resistance vs. gate length $L_g$ for FETs with regrown InAs contacts. Inset: Completed FET with Ti/Au source/drain pads and $L_g$=1 μm gate finger

Figure 6. Surface mobility affects facet competition. Planes with high surface mobility suffer a net loss of atoms to planes with stronger bonds. A-C correspond to conditions observed in Fig. 3.

Figure 7. Gap mechanism. A: Weak adhesion of In and Ga on $SiO_2$ leads to migration to neighboring semiconductor surface. B: Arriving Ga/In atoms locally enrich the III/V ratio, leading to rapid surface diffusion (C). As III/V ratio decreases farther from gate, growth begins on low-index facets (D) and eventually becomes planar and As-terminated (E).